\documentclass[useAMS]{mn2e}
\usepackage{times}
\usepackage{graphicx}

\pubyear{2005}

\author[Ray, Bagla and Padmanabhan]
{Suryadeep Ray$^1$, J.S. Bagla$^1$ and T. Padmanabhan$^2$ \\
  $^1$ Harish-Chandra Research Institute,  Chhatnag Road, Jhunsi, Allahabad
  211019, India. \\
  $^2$ Inter-University Centre for Astronomy and Astrophysics, Post Bag 4,
  Ganeshkhind, Pune 411007, India. \\
  E-mail: surya@mri.ernet.in, jasjeet@mri.ernet.in, nabhan@iucaa.ernet.in} 

\title[Gravitational Collapse in an Expanding Universe]
{Gravitational Collapse in an Expanding Universe : Scaling Relations for
Two-Dimensional Collapse Revisited}

\label{firstpage}

\def\LaTeX{L\kern-.36em\raise.3ex\hbox{a}\kern-.15em
    T\kern-.1667em\lower.7ex\hbox{E}\kern-.125emX}

\pagerange{\pageref{firstpage}--\pageref{lastpage}} \pubyear{2005}

\begin{document}

\maketitle

\begin{abstract}
We investigate non-linear scaling relations for two-dimensional gravitational
collapse in an expanding background using a 2D TreePM code and study the
strongly non-linear regime ($\bar\xi \leq 200$) for power law models.  
Evolution of these models is found to be scale-invariant in all our
simulations. 
We find that the stable clustering limit is not reached, but there is a model
independent non-linear scaling relation in the asymptotic regime.
This confirms results from an earlier study which  only probed the
mildly non-linear regime ($\bar\xi \leq 40$). 
The correlation function in the extremely non-linear regime  is
a less steep function of scale than reported in earlier studies. 
We show that this is due to coherent transverse motions in massive haloes. 
We also study density profiles and find that the scatter in the inner and
outer slopes is large and that there is no single universal profile that fits
all cases. 
We find that the difference in typical density profiles for different models
is smaller than expected from similarity solutions for halo profiles and
transverse motions induced by substructure are a likely reason for this
difference being small. 
\end{abstract}




\section{Introduction}

Large-scale structures such as clusters of galaxies are very
over-dense compared to the average density of the universe. 
It is believed that these were formed by the growth of small perturbations via 
gravitational instability
\cite{2002PhR...367....1B,2002tagc.book.....P,1999coph.book.....P,1980lssu.book.....P}.  
In general, the equations that describe the growth of density perturbations
due to gravitational clustering \cite{1974ApJ...189L..51P,1980lssu.book.....P}
cannot be solved analytically when the over-densities are large and one has to
rely on N-Body simulations for detailed predictions.  
Limited information about non-linear gravitational clustering can be obtained
by using approximations or ansatze, e.g., stable clustering
\cite{1980lssu.book.....P} for which it is assumed that the infall velocity
and Hubble velocity compensate for each other leading to a stable
configuration.    
Stable clustering allows us to relate the initial spectrum of fluctuations with
the final asymptotic spectrum of fluctuations \cite{1977ApJS...34..425D}. 
Non-linear scaling relations are a more detailed prescription for relating the
initial and final spectrum of fluctuations
\cite{1991ApJ...374L...1H,1994MNRAS.271..976N,1994MNRAS.267.1020P,1995MNRAS.276L..25J,1996MNRAS.280L..19P,1996MNRAS.278L..29P,1996ApJ...466..604P,1997MNRAS.286.1023B,1998ApJ...495...25B,2003MNRAS.341.1311S}. 
The motivation for studying non-linear scaling relations is to understand key
phases in gravitational clustering, and to identify the relevant process in
each phase \cite{1996MNRAS.278L..29P}. 

Existence of non-linear scaling relations implies that gravitational
clustering does not erase memory of initial conditions.  
This provides another useful and a more prosaic motivation for studying
scaling relations, namely to be able to make predictions about clustering of
galaxies for a given power spectrum \cite{1991ApJ...374L...1H}. 
Scaling relations can also be used to make predictions for weak lensing
observations in a given model \cite{1992ApJ...388..272K,1997ApJ...484..560J}. 
These scaling relations are valid for hierarchical models where the initial 
conditions contain fluctuations at all scales and the amplitude of
fluctuations increases monotonically as we go from large-scales to
small-scales.  

Non-linear scaling relations indicate that there are three prominent regimes 
in evolution of gravitational clustering.
When the amplitude of perturbations is small so that the density contrast
($\delta = \rho/{\bar\rho} - 1$) is close to zero ($|\delta| \ll 1$), mode
coupling is not important and the evolution closely follows the predictions of
linear perturbation theory \cite{1974ApJ...189L..51P}.  
The power spectrum and correlation function evolve without a change in shape
in the linear regime. 
As the density contrast grows and becomes comparable to unity, motions induced
by gravitational collapse start to dominate over the expansion of the
universe.   
The quasi-linear regime is dominated by infall onto density peaks and the
correlation function grows rapidly in this phase \cite{1996MNRAS.278L..29P}. 
Gravitational collapse leads to formation of structures in or close to
dynamical equilibrium and further evolution of density contrast is dominated by
depletion of average density due to expansion of the universe as the density
of collapsed structures remains almost constant.  
We shall call this the asymptotic regime. 

Initial conditions with power law power spectra are very useful for such
studies.  
These power spectra have the form $P(k) \propto k^n$ with $n$ being the index
of the power spectrum.  
We require $(n+D) > 0$ for hierarchical clustering and N-Body simulations
can be used for only this type of models, indeed even for these models there
are restrictions on the size of the simulation box in order for the simulation
to reproduce the model correctly \cite{2004astro.ph.10373B}.
Here $D$ is the number of spatial dimensions in which gravitational clustering
is being studied. 
If we consider an Einstein-deSitter universe then such a model is
characterised by the scale of non-linearity $r_{NL}$ and the index $n$, where
$r_{NL}$ is defined as the scale at which rms fluctuations in linearly
extrapolated density contrast ($\sigma_L(r)$) is unity. 
\begin{eqnarray}
\sigma_L^2(r) ~ &=& ~ a^2 ~ r^{-\left(n+D\right)} \nonumber \\
r_{NL} ~ &=& ~ a^{2/\left(n+D\right)}
\end{eqnarray}
If the evolution of clustering is self-similar then the non-linear correlation
function plotted as a function of $r/r_{NL}$ should have the same shape at all
epochs for a given index $n$. 
Note that the evolution is expected to be self-similar if $r_{NL}$ is the only
scale in the problem and it is not expected in cosmologies with non-zero
spatial curvature or a non-power-law expansion factor.

Several studies have attempted to understand the nature of the asymptotic
regime, mainly with help of N-Body simulations
\cite{1994MNRAS.267.1020P,1995MNRAS.276L..25J,1996ApJ...465...14C,1996ApJ...466..604P,1996MNRAS.280L..19P,1997MNRAS.287..687J,2000ApJ...531...17K,2001ApJ...550L.125J,2003MNRAS.341.1311S}. 
Key conclusions of these studies about the asymptotic regime may be summarised
as follows. 
\begin{itemize}
\item
Stable clustering is not reached in the range of non-linearities explored by
N-Body simulations. 
However, the departure from stable clustering for most models is small.
\item
Gravitational clustering does not erase memory of initial conditions, i.e.,
the non-linear power spectrum is not independent of the linear power
spectrum to the extent it could be probed.
There is no {\it universal} asymptotic regime.
\end{itemize}
One of the reasons for the inability to resolve the problem of asymptotic
regime has been the limited dynamic range of N-Body simulations.
The problem of dynamic range can be circumvented by simulating a
two-dimensional system
\cite{1985A&A...144..413B,1999ApJ...512....1E,2000PhRvD..61b3515P} instead of
a three-dimensional one, wherein a much higher dynamic range can be achieved
with similar computational resources
\cite{1998ApJ...495...25B,1998MNRAS.293L..68M}.  
Generic features like non-linear scaling relations are likely to be
independent of dimension and --- in fact --- the scaling relations in the
quasi-linear regime were predicted \cite{1996MNRAS.278L..29P} well before
these could be tested in simulations \cite{1998ApJ...495...25B}. 
If we can understand the nature of the asymptotic regime in two dimensions, it
will help us solve the problem in three dimensions even if we cannot map the
solution directly to the full problem in three dimensions.
Previous studies of gravitational clustering in two dimensions concluded that 
there is no stable clustering \cite{1998ApJ...495...25B,1998MNRAS.293L..68M}. 
Both the studies mentioned above were limited to $\bar\xi \leq 40$ and the
dynamic range in the non-linear regime was limited compared to the present
study (see eqn.(6) for the definition of $\bar\xi$).
In this work we revisit the same issues, but using the 2D TreePM code
\cite{2004astro.ph..6009R} for N-body simulations. 
The TreePM code has a better force resolution as compared to a PM code
\cite{1985A&A...144..413B}, therefore in our simulations we have a
significantly larger dynamic range over which we can study the asymptotic
regime. 

A related issue is that of density profiles of haloes surrounding galaxies and
clusters. 
Non-linear scaling relations in the asymptotic regime depend on the kind of
dynamical equilibrium that is reached in massive haloes, if an equilibrium
is ever reached. 
Under the assumption of self-similar collapse, the density profiles of massive
haloes can be related to the initial conditions, and also to the dynamics
within these haloes \cite{1984ApJ...281....1F,2000ApJ...538..517S}. 
If these models are true then the slope of density profile at small scales is
related to the index of power spectrum as:
\begin{equation}
\rho(r) \propto r^{-\eta} ~~~~~ ; ~~~ \eta ~ = ~ \frac{D ~ \left( n + D +
  1 \right)}{2 + D + n} 
\end{equation}
If self-similar collapse of haloes is the correct description then the final
density profile of a halo carries information about the initial profile and
hence about the power spectrum \cite{1986ApJ...304...15B}.
It has been claimed that gravitational clustering in hierarchical models leads
to a {\it universal} density profile \cite{1996ApJ...462..563N} irrespective
of the initial conditions, we shall refer to this density profile as the NFW
profile. 
\begin{equation}
\rho_{_{\rm NFW}}(r) \propto \frac{1}{(r/r_s) ( 1 + r/r_s)^2} 
\label{NFW}
\end{equation}
Here $r_{s}$ is a scale radius. 
We will also use the term $r_{200}$ for this.
This profile is characterised by a $r^{-3}$ decline at large radii and a cuspy
inner profile of the form $\rho(r) \propto 1/r$.
The claim of universality has been tested in several studies
\cite{1996ApJ...462..563N,1996MNRAS.281..716C,1997ApJ...477L...9F,1997MNRAS.286..865T,1998ApJ...499L...5M,1998ApJ...502...48K,2000ApJ...544..616G,2000ApJ...538..528S,2001ApJ...554..903K,2003MNRAS.338...14P,2004ApJ...606..625F,2004MNRAS.349.1039N}
and though there are disagreements, it appears that the NFW profile is
{\it consistent} with the density profiles of massive haloes in N-Body
simulations even if it is not the best fit. 
This essentially implies that there is a scatter in density profiles and
one can fit different functional forms to the N-Body data
\cite{2000ApJ...535...30J,2001MNRAS.321..559B}.   

If true, the existence of NFW like profile implies that there are some
universal aspects of gravitational clustering in an expanding universe. 
There is no {\it ab initio} derivation of the NFW profile, but one can argue
that if there is a universal profile it should have same asymptotes as the
form shown above
\cite{1998MNRAS.293..337S,1999MNRAS.303..685N,2002LNP...602..165P}.  

We briefly review non-linear scaling relations in the following section.  
Numerical simulations used in the present work are discussed in \S{3}. 
We conclude in \S{4}.


\section{Non-linear Scaling Relations}

In this section we review the non-linear scaling relations for gravitational
clustering in two dimensions. 
The non-linear and the linear correlation functions at two different
scales can be related by non-linear scaling relations
\cite{1991ApJ...374L...1H,1994MNRAS.271..976N}.  
The relation between these scales is given by the characteristics of the pair
conservation equation \cite{1980lssu.book.....P,1994MNRAS.271..976N} under the
assumption that the pair velocity function $h$ (see below) depends on $r$ and
$t$ through $\bar\xi$ \cite{1994MNRAS.271..976N}. 
For the two dimensional system of interest, this equation can be written as
\begin{equation}
{\partial D \over \partial A} -h (A, x) {\partial D \over \partial X}
= 2h (A, X) .
\end{equation}
Here 
\begin{equation}
D=\log(1 + \bar\xi),\quad X=\log(x),\quad A=\log(a),
\end{equation}
and
$h = - v_p/\dot a x$ is the ratio between mean pair
velocity and Hubble velocity. 
The volume averaged correlation function is defined as:
\begin{equation}
\bar\xi(x)=2 x^{-2} \int^x r \xi(r) dr
\end{equation}
where $\xi$ is the correlation function.
The characteristics of this equation are $x^2 (1 + \bar \xi(x,a)) = l^2 $,
where $x$ and $l$ are the two scales used in non-linear scaling relations.  
The self-similar models \cite{1984ApJ...281....1F} imply that for collapse of
cylindrical perturbations the turn-around radius and the initial density
contrast inside that shell are related as $x_{\rm ta} \propto
l/\bar\delta_i \propto l/\bar\xi_L (l)$.  
Here $\bar\xi_L$ is the linearly extrapolated mean correlation function.  
Noting that in two dimensions mass enclosed in a shell $M \propto x^2$, we
find $\bar\xi (x) \propto \left[ \bar\xi_L (l) \right]^2$ in the regime
dominated by infall.   
Thus in 2-D the scaling relations are
\begin{equation}
\bar \xi (a,x) \propto \left\{ \hbox{  }
\begin{array}{ll} 
\bar \xi_L (a,l) & \hbox{ (Linear)} \\
\bar \xi_L(a,l)^2  & \hbox{ (Radial Infall)} \\
\bar \xi_L(a,l)^h &  \hbox{ (Asymptotic limit)} \\
\end{array} \right.
\label{hamilton}
\end{equation}
The usual stable clustering limit is $h=1$ where the infall and Hubble
expansion balance each other in structures in dynamical equilibrium.  
In such a case, or for any value of $h$ that does not depend on the initial
conditions, slope of the non-linear correlation function is a unique function
of the slope of the initial correlation function. 
For a general $h$ we can relate the slope of the correlation function in the
asymptotic regime to the slope of the initial linear correlation function 
\begin{equation}
\bar \xi (a,x) \propto x^{-2h\left( n+2 \right) / \left( 2 + h \left( n+2
    \right) \right)} .
\end{equation}
If, however, $h(n+2) = $~constant then the non-linear correlation function has
the same slope {\it independent of the initial correlation function.}
This will happen if gravitational clustering erases all memory of the initial
conditions. 
Note that this will make $h$, and hence the velocity fields a function of the 
initial conditions.

In our earlier work we had concluded that $h \simeq 0.75$ in the asymptotic
regime \cite{1998ApJ...495...25B}.
In the same work we confirmed the prediction for the infall dominated
quasi-linear regime, we found that the index in this regime is indeed $2$. 
(Note that in 3D, the indices for three regimes are $1$, $3$ and $3h/2$
respectively.) 
In this work, we shall, among other things address the following two
questions: 
\begin{itemize}
\item
Does the asymptotic value of $h$ scales with $n$ such that $h (n + 2 )=$~\rm
constant? 
If it does, then the final slope of the non-linear correlation function will
be independent of the initial slope. 
We will show that this does not happen. 
\item
Does $h$ reach a universal value independent of $n$? 
This seems to be the case and we will show that the behaviour of $h$ is
independent of $n$. 
But since all simulations, including this one probes only up to finite $\xi$,
its actual asymptotic value is difficult to determine. 
\end{itemize}


\section{Numerical Simulations}

Our aim in this paper is to study the nature of scaling relations in the
asymptotic regime.  
We choose to work in two dimensions in order to improve the dynamic range over
which the results of N-Body simulations are reliable. 
We wish to do this without changing the system completely as the ultimate goal
is to carry over the understanding to three dimensional systems. 
Therefore we study the evolution of two dimensional perturbations in a three
dimensional expanding universe \cite{1984ApJ...281....1F}. 
We start with a set of infinitely long straight ``needles'' all pointing along
one axis.  
The evolution keeps the ``needles'' pointed in the same direction and we study
gravitational clustering of these needles in an orthogonal plane. 
Particles in the N-body simulation represent the intersection of these
``needles'' with this plane and the force of interaction between ``particles''
is given by the solution of the Poisson equation in two dimensions. 
The inter-particle force falls as $1 / r$, unlike the inverse square force in
three dimensions. 

We use the two dimensional TreePM code \cite{2004astro.ph..6009R}, a modified
version of the three dimensional TreePM code
\cite{2002JApA...23..185B,2003NewA....8..665B}. 
The TreePM code splits the interaction force into two parts, the long range
force is computed in Fourier space and the short range component is computed
in real space.  
The TreePM code does not use the usual PM force as the long range force, this
allows us to control errors in force over the entire range of scales
\cite{2002JApA...23..185B,2003NewA....8..665B,2004astro.ph..6009R}. 

We soften the two-dimensional gravitational force at small scales in order to
ensure {\it collisionless evolution} of the particle distribution in our
simulations. 
We use a cubic spline form for softening, i.e., each particle is assumed to
have an extended mass distribution represented by the normalised spline kernel
used in the SPH formalism \cite{1992ARA&A..30..543M}.
This softens the force at scales smaller than the softening length $\epsilon$,
and at scales larger than $\epsilon$ we get the full $1 / r$ force. 
All N-Body simulations reported in this paper used $\epsilon \geq 0.2$ in
units of the average inter-particle separation. 

We have carried out a series of numerical simulations with power law initial
spectra with indices $n=-0.4$, $0.0$ and $1.0$. 
Several realisations of each power spectrum were simulated.  
Scatter in values across this suite of simulations was used to estimate the
error bars.  
It is difficult to reach the asymptotic regime for $n \ll -0.4$ before
the perturbations at the box scale become important and hence we have not used
models with a more negative index. 
Models were normalised so that $\Delta^2(k=8k_f,a=1) = 1$ where $k_f$ is the
wave number of the fundamental mode, essentially the fluctuations at $1/8$ of
the box size were normalised to unity at $a=1$.
Simulations were run upto a sufficiently late epoch while requiring that the
fluctuations at half the box scale were well within the linear regime at the
final epoch, i.e., $\Delta^2(k=2k_f,a_{fin}) \ll 1$.  

Simulations discussed here used $2048^2$ particles on a $2048^2$ grid in an
Einstein de Sitter background universe. 
TreePM code was used for all the simulations, though we have checked many of
the results in larger $4096^2$ PM simulations.
We used $r_s=1.0$ grid length in the TreePM code, this is the scale used for
dividing the force into a short range and a long range component
\cite{2002JApA...23..185B,2003NewA....8..665B,2004astro.ph..6009R}. 
Error in force depends largely on the choice of this scale and the cell
opening criterion, we use $\theta_c=0.5$.

In an Einstein-deSitter universe the evolution of perturbations with a power
law power spectrum is expected to be scale-invariant.  
The only scale introduced by gravitational clustering is the scale of
non-linearity $r_{NL}$.  


\begin{figure}
\includegraphics[width=3.2truein]{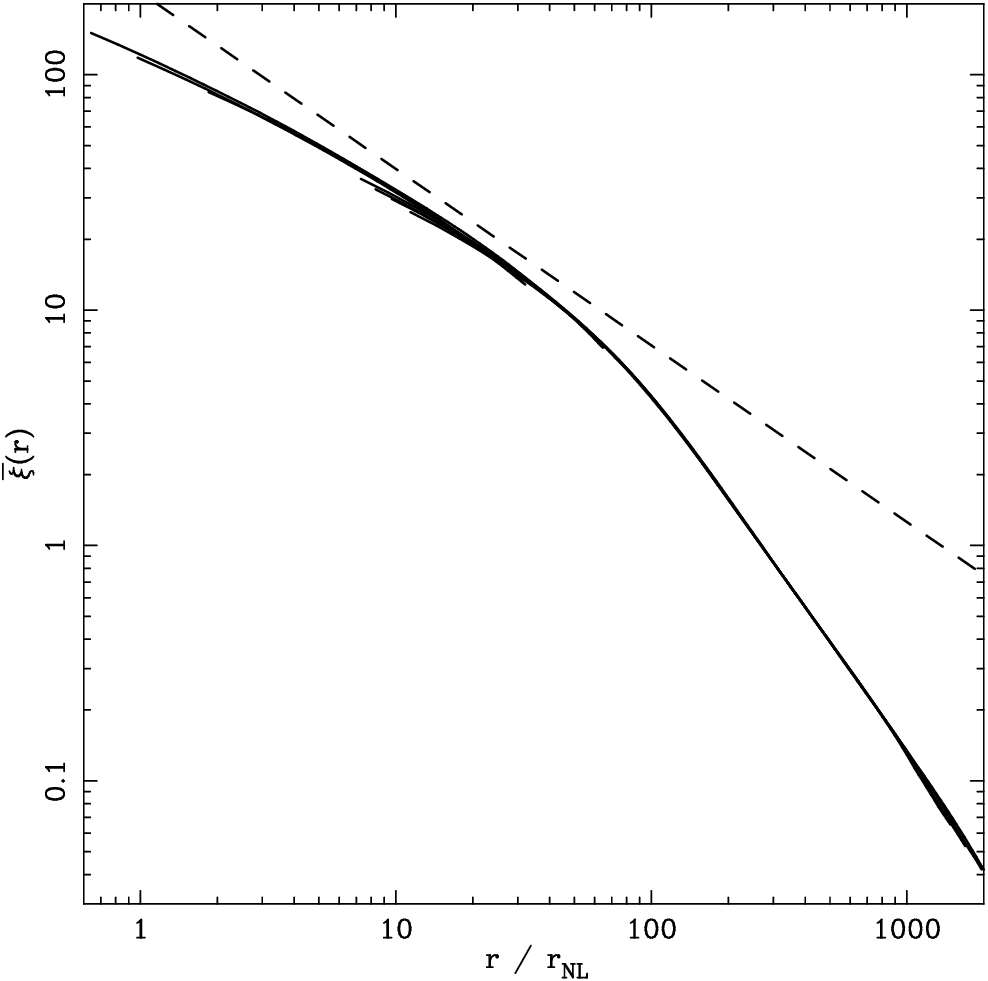} 
\includegraphics[width=3.2truein]{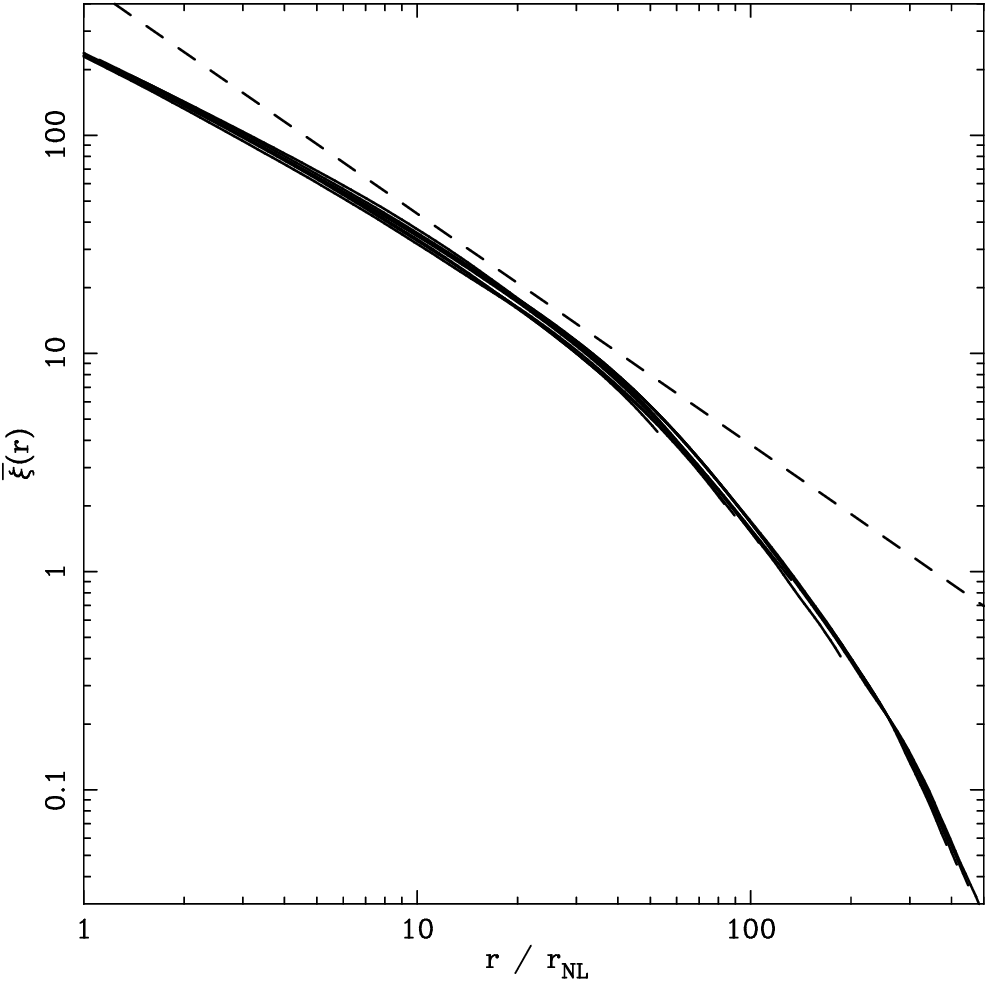}
\caption{$\bar\xi$ is plotted as a function of $r / r_{_{\rm NL}}$ for
  $n=-0.4$ (upper panel) and $n=1.0$ (lower panel) simulations.  These
  curves span more than a decade in scale factor.  Clearly, over the
  entire range of clustering the evolution is self-similar.  Dashed lines
  in these panels shows the asymptotic slope for $h=0.75$, a result
  suggested by an earlier study.} 
\label{fig_corr}
\end{figure}


We outline the key results of these simulation studies in the
following subsections. 


\subsection{Correlation Function}

We find that the evolution of the system is self-similar in that all
the relevant quantities have the same form when scaled by
$r_{_{\rm NL}}\propto a^{2/(n+2)}$, where $r_{_{\rm NL}}$ is the scale where
the variance of density fluctuations is unity.
Fig.~\ref{fig_corr} shows the averaged correlation function
$\bar\xi(r)$ as a function of $r / r_{_{\rm NL}}$ for simulations with
$n=-0.4$ (upper panel) and $n=1.0$ (lower panel). 
Curves were plotted for scales larger than twice the softening length
in order to keep out features that depend on the choice of softening
length. 
The evolution is self-similar up to and beyond $\bar\xi = 100$. 

The shape of the correlation function $\bar\xi(r)$ for weaker
non-linearities is consistent with the results of earlier studies
\cite{1998ApJ...495...25B,1998MNRAS.293L..68M}. 
There was disagreement between the two studies for the asymptotic
regime: one study \cite{1998ApJ...495...25B} favoured $h \simeq 0.75$ and the
other found $h (n+2)=1.3$ \cite{1998MNRAS.293L..68M}. 
The latter behaviour erases memory of initial conditions and the slope
of the correlation function in the asymptotic regime is the same for all 
models. 
The dynamic range in both the studies was limited to $\bar\xi \sim 10$. 

\begin{figure}
\includegraphics[width=3.2truein]{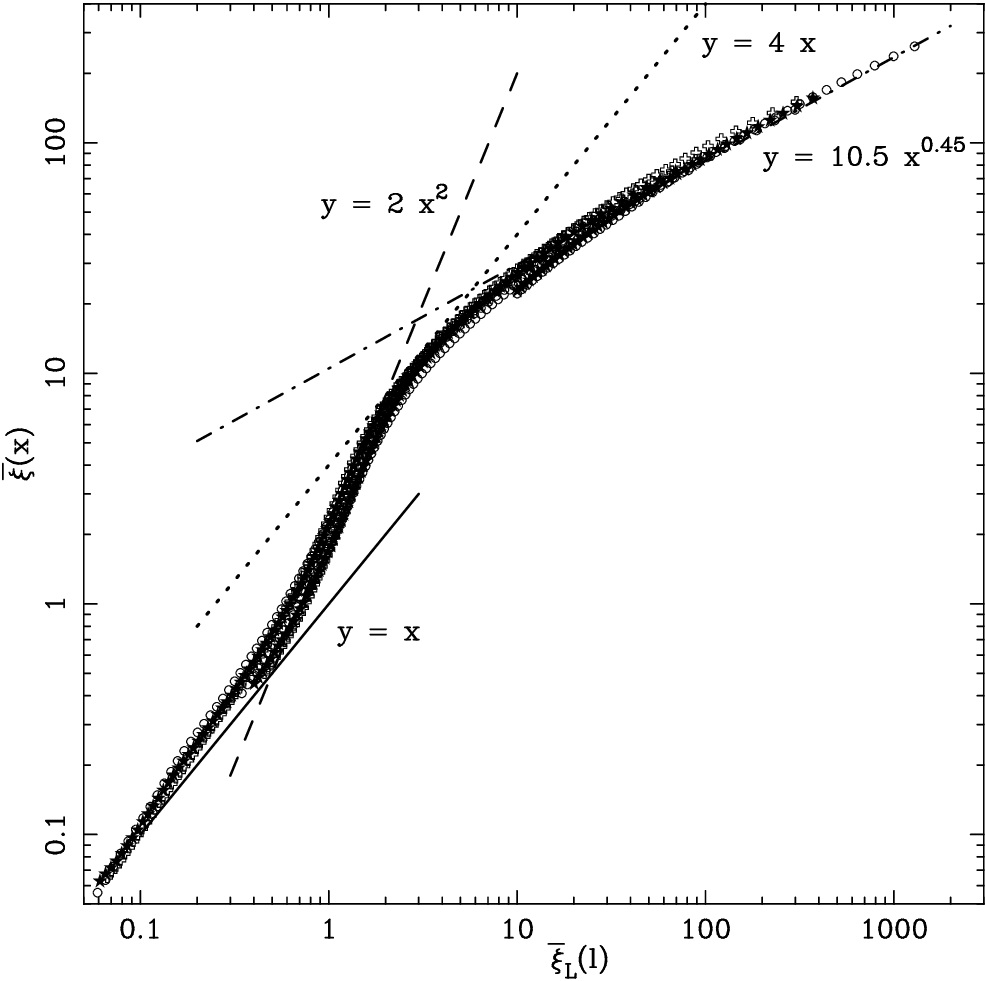}
\includegraphics[width=3.2truein]{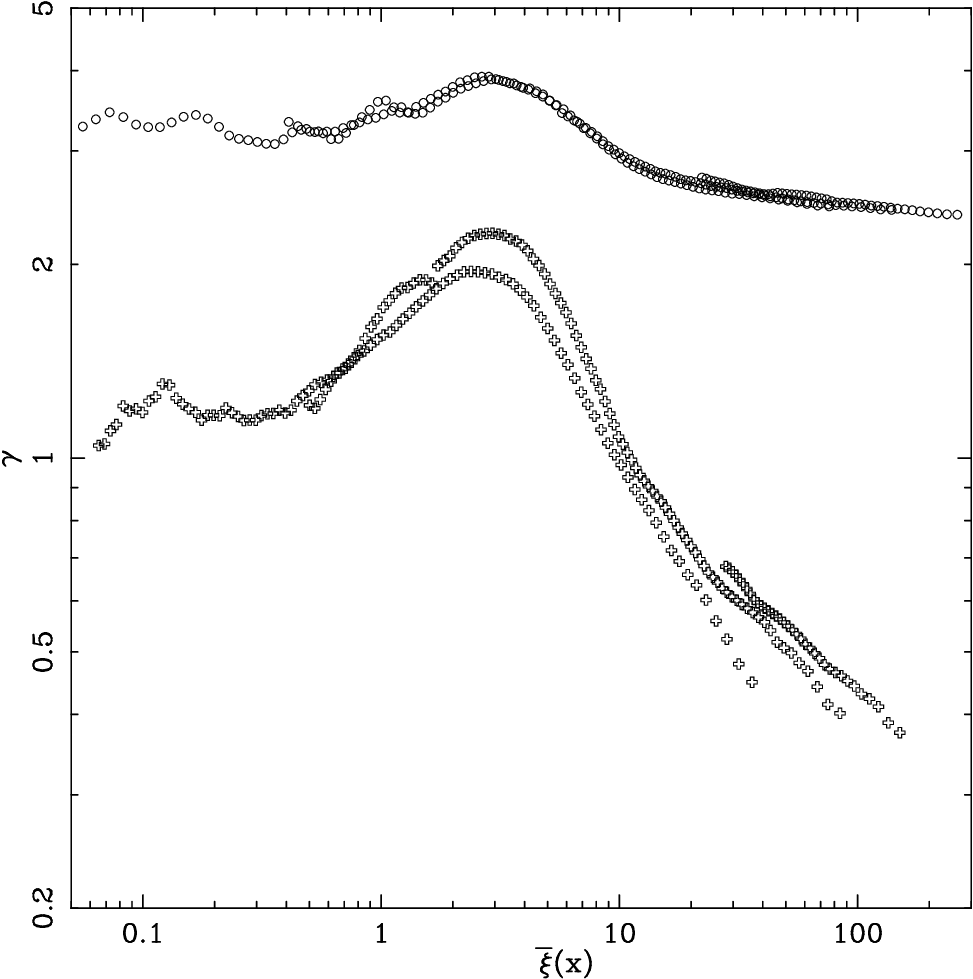}
\caption{The upper panel shows the non-linear scaling relation.
  $\bar\xi(x,a)$ is plotted as a function of $\bar\xi_L(l,a)$ for
  $n=-0.4$, $n=0$ and $n=1$.  Data from multiple epochs has been used
  here.  Curves marking the linear (thick curve), quasi-linear (dashed
  curve) and asymptotic regime (dot-dashed curve) are shown here.  We
  have also marked a dotted line showing the stable clustering limit.
  A remarkable feature of clustering in two dimensions is that the
  non-linear correlation function in the asymptotic regime is {\it
  smaller} than the linear correlation function.  The slope $\gamma$ is shown
  as a function of $\bar\xi(x,a)$ in the lower panel.  Circles mark the slope
  for the $n=1$ model and the plus sign marks the slope for the $n=-0.4$
  model.  Points for the $n=1$ model are displaced upwards by $2$ for
  clarity.}   
\label{fig_nsr}
\end{figure}

The current work with higher dynamic range agrees with the  
slope for $h = 0.75$ in the region of overlap with the previous work 
\cite{1998ApJ...495...25B} and {\it rules out} the possibility of $h
(n+2)=1.3$. 
The dashed line with the expected slope for $h = 0.75$ is marked in
Fig.~\ref{fig_corr} and it runs parallel to the curve up to about
$\bar\xi \simeq 40$, the largest non-linearity studied earlier
\cite{1998ApJ...495...25B}. 
As clustering increases, the slope of the correlation function
decreases below the slope expected for $h = 0.75$.  
Slope of the correlation function for these two models is shown in
Fig.~\ref{fig_nsr}. 
The asymptotic slope of the correlation function is $-0.53 \leq \gamma \leq
-0.50$ for $n=-0.4$ and $-0.80 \leq \gamma \leq -0.77 $ for $n=1$,
where $\gamma = \partial\log\bar\xi / \partial\log r$ and is evaluated
at $\bar\xi \geq 100$.  
We have given the $95\%$ confidence limits for the slope, the limits were
derived using a $\chi^2$ fit to the correlation function in the
asymptotic regime. 
Different values of $\gamma$ imply departure from the $h (n+2) =
$~constant asymptote.
The values of $\gamma$ listed above are consistent with $0.416 \leq h
\leq 0.451$ for $n=-0.4$ and $0.417 \leq h \leq 0.444$ for $n=1$.  
Thus we recover a similar range of asymptotic values for $h$ from
the correlation function. 
As we will show in \S{3.2}, the slope of the correlation function is a more
stable estimator of $h$ than a direct determination. 


Fig.~\ref{fig_nsr} shows the non-linear scaling relations.  
$\bar\xi(x,a)$ is plotted as a function of $\bar\xi_L(l,a)$ for
$n=-0.4$, $n=0$ and $n=1$.  
Data from multiple epochs have been used here.  
Curves marking the linear, quasi-linear and asymptotic regime are
shown here.  
We have also marked a line showing the stable clustering limit.
The equations for the piecewise fit are:
\begin{equation}
\bar \xi (a,x) = \left\{ \hbox{  }
\begin{array}{ll} 
\bar \xi_L (a,l); & \hbox{$\bar\xi(x)\leq 0.5$}\\
 2 \,\, {\bar\xi_L(a,l)}^2;  & \hbox{$0.5 \leq
\bar\xi(x)\leq 8 $} \\
 10.5 \,\, {\bar\xi_L(a,l)}^{0.45}; &  \hbox{$22 \leq 
\bar\xi(x)$ } \\
\end{array} \right .
\label{ourfit}
\end{equation}
It is remarkable that the points for all the three models follow the
same non-linear scaling relation.  
The asymptotic slope implied by the scaling relation lies near
$h=0.45$ but it is difficult to say whether this is the final value or
if this continues to decrease as we go to higher non-linearities. 
This evolution is inconsistent with the asymptote of $h(n+2)=$~constant, 
as the scaling relations for different models would have been
different in that case.

The results shown here with the TreePM code match with the correlation
function obtained in $4096^2$ PM simulations. 
This comparison has been carried out up to $\bar\xi \leq 70$.


\begin{figure}
\includegraphics[width=3.2truein]{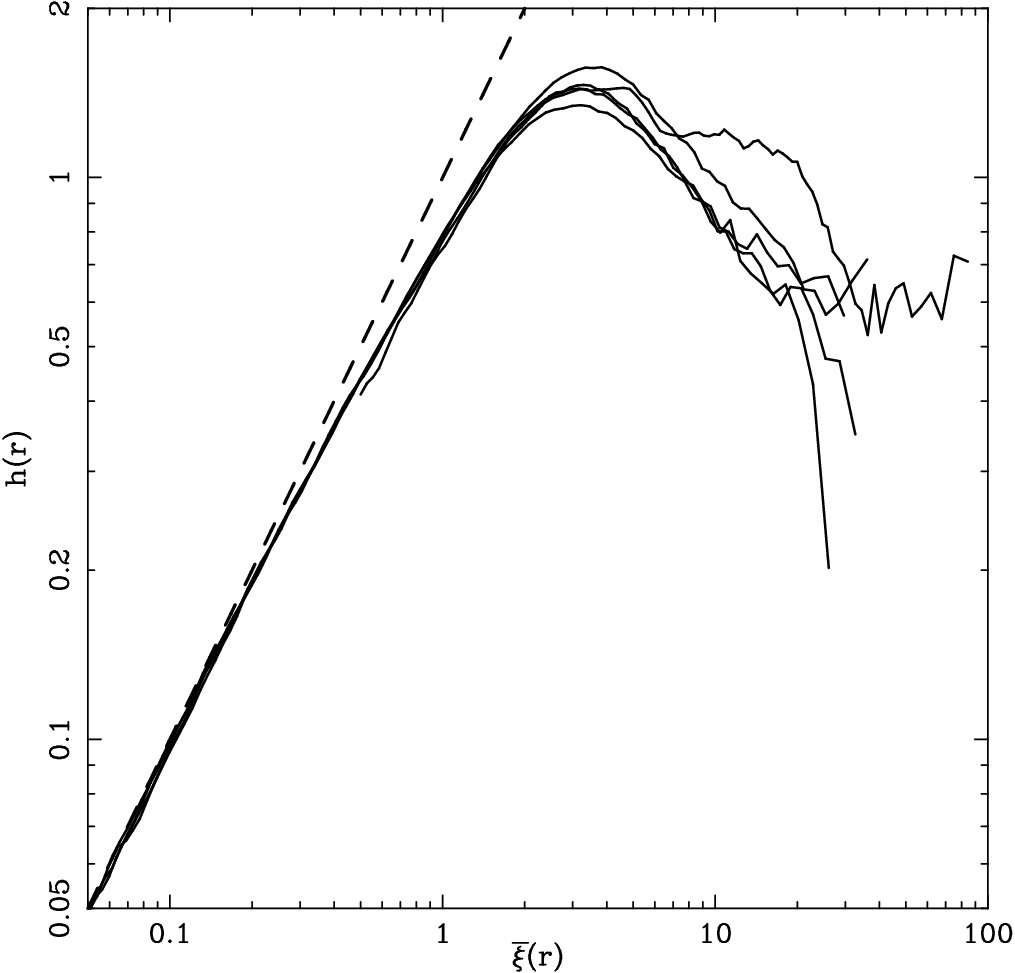} 
\includegraphics[width=3.2truein]{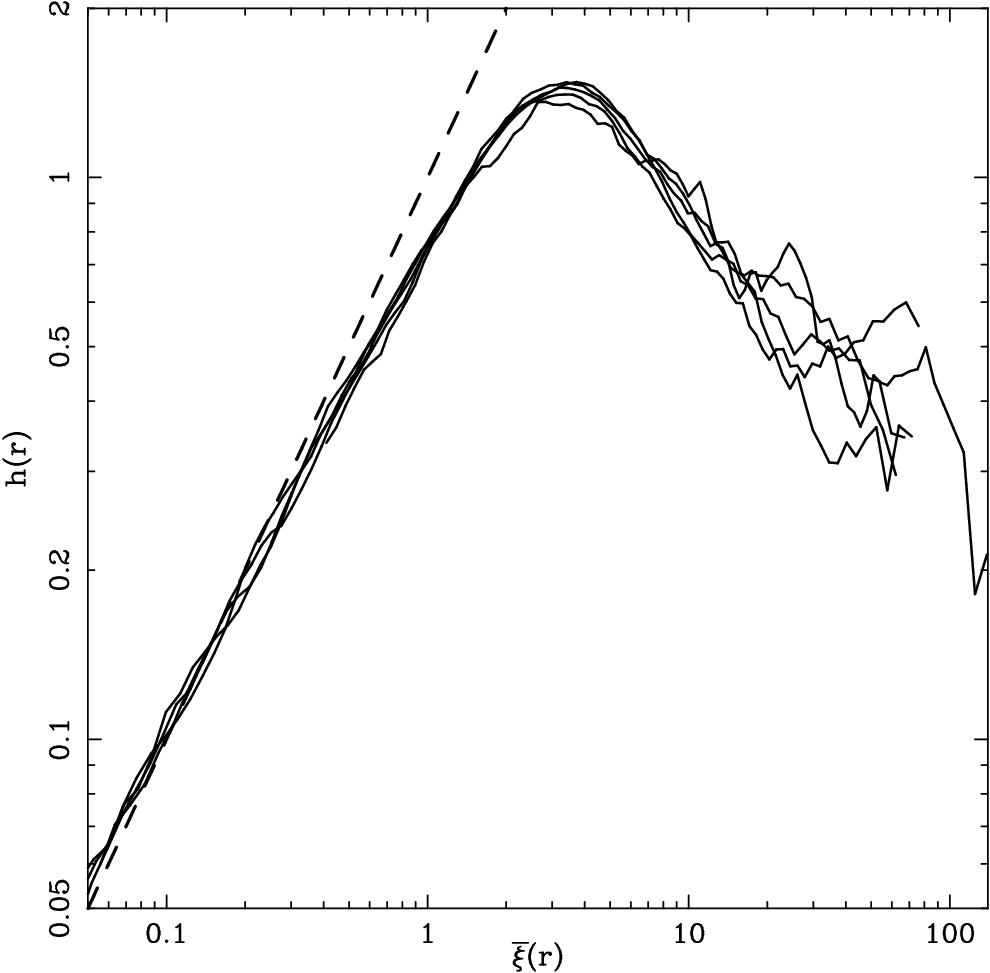}
\caption{The pair velocity $h(r,a)$ is plotted as a function of
  $\bar\xi(r,a)$ for a large number of epochs $a$ for $n=-0.4$ (upper
  panel) and $n=1.0$ (lower panel).  The dashed line shows the
  expected value of $h$ in the linear limit.  The $h-\bar\xi$ curves are
  the same across all epochs and for all models within the scatter.}
\label{fig_pairv}
\end{figure}


\begin{figure}
\includegraphics[width=3.2truein]{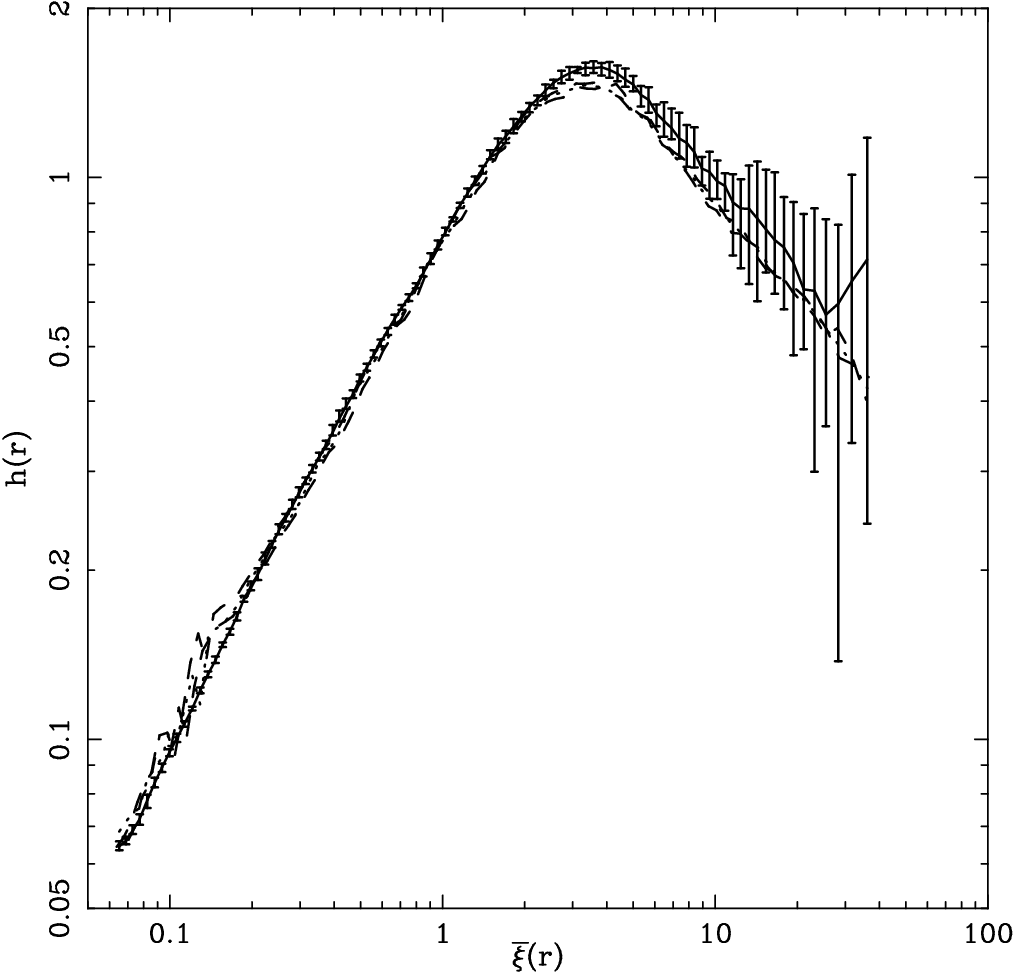} 
\caption{The pair velocity $h(r,a)$ is plotted as a function of
  $\bar\xi(r,a)$ at late times for $n=-0.4$.  Error bars are also shown here.
 The range of $h(r,a)$ expected from eqn.(4) is shown here using dashed
 lines.  The expected values and the directly measured values of $h$ are
 consistent within the error bars for nearly the entire range of scales.}
\label{fig_pairv_nsr}
\end{figure}


\begin{figure}
\includegraphics[width=3.2truein]{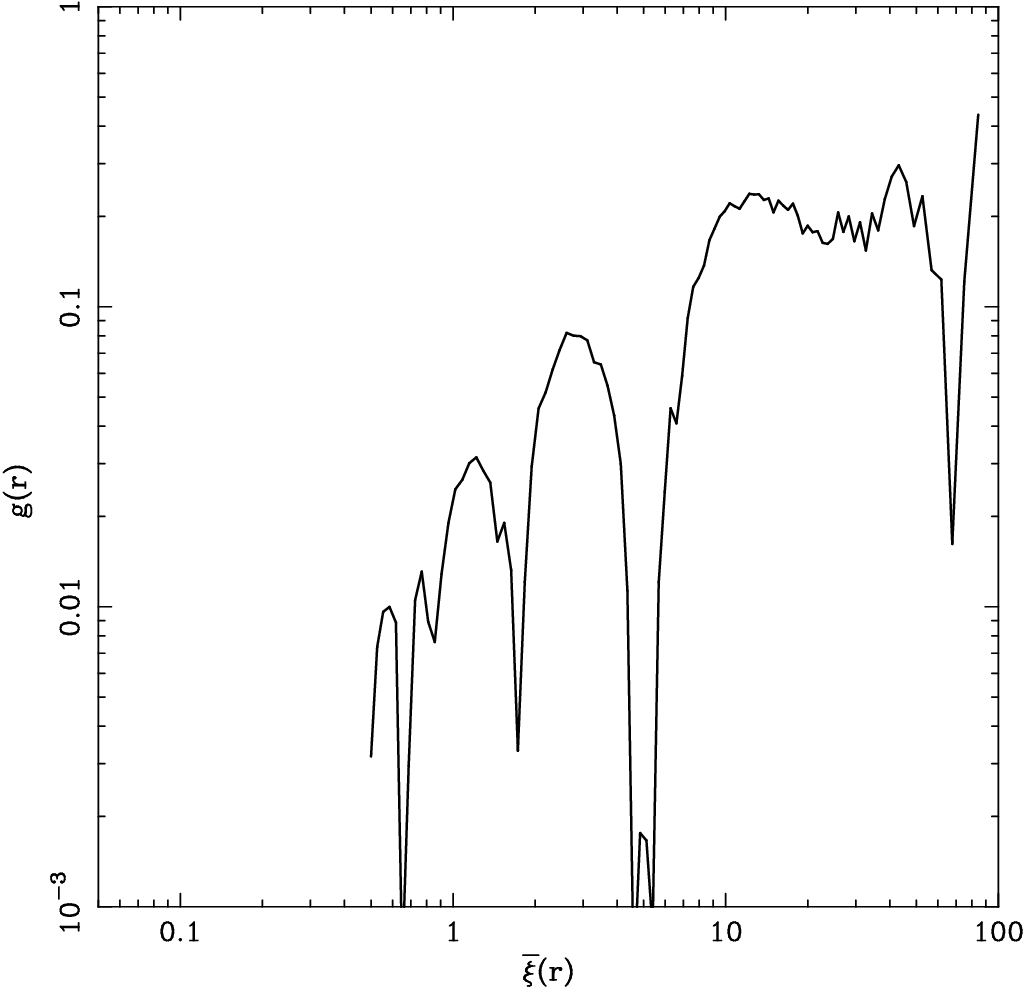} 
\includegraphics[width=3.2truein]{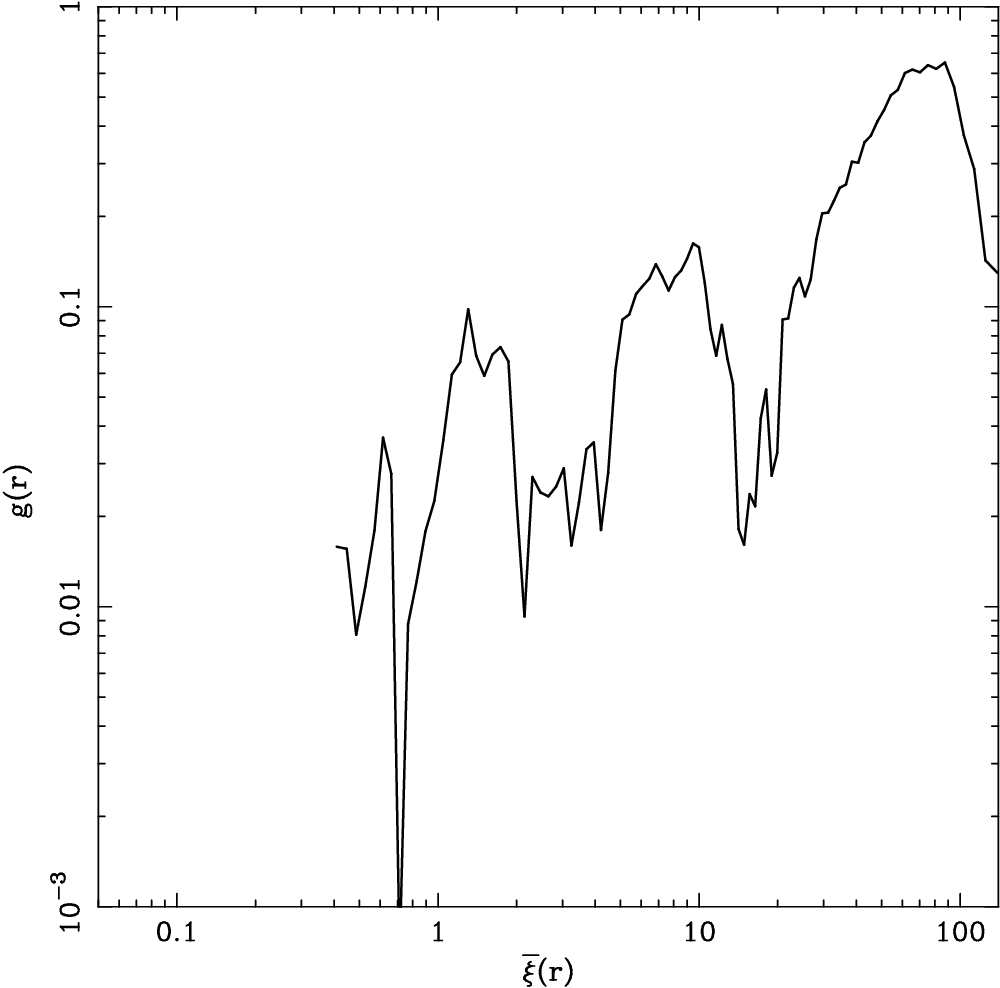}
\caption{The transverse component of pair velocity $g(r,a)$ is plotted as a
  function of $\bar\xi(r,a)$ for a late epoch for $n=-0.4$ (upper
  panel) and $n=1.0$ (lower panel) models.  The magnitude $|g|$ is plotted
  here, a remarkable fact apparent from this figure is that the transverse
  component is more important for $n=1$.  At $\bar\xi \gg 1$, the magnitude of
  $g$ is comparable with that of $h$.}
\label{fig_pairvt}
\end{figure}


\subsection{Pair Velocity}

Fig.~\ref{fig_pairv} shows the pair velocity $h(r,a)$ as a
function of $\bar\xi(r,a)$ for a large number of epochs $a$ for
$n=-0.4$ (upper panel) and $n=1.0$ (lower panel).  
The dashed line shows the expected value of $h$ in the linear limit.
For the linear and quasi-linear regime ($\bar\xi \leq 10$) $h$ is a single 
valued function of $\bar\xi$, irrespective of the epoch for a given
model. 
Indeed, we find that there is no significant difference between the
curves for different models at $\bar\xi \leq 10$, and at higher
non-linearities there is a large scatter in the curves so it is
difficult to test any claims using Fig.~\ref{fig_pairv}.
Thus simulations are consistent with the ansatz that $h$ depends on epoch and
scale {\it through} $\bar\xi$ \cite{1994MNRAS.271..976N} and this allows us to
find the form of the non-linear scaling relation.  
We can certainly do this for a given model and as we do not find significant
differences between different models, there does not seem to be any power
spectrum dependence in scaling relations. 
However, there will be a dependence on cosmology \cite{1996ApJ...466..604P}.
Conversely, non-linear scaling relations can be used to find
$h(\bar\xi)$. 

We have checked that the scaling relations plotted in
Fig.~\ref{fig_nsr} are consistent with the $h(\bar\xi)$ curves
plotted in Fig.~\ref{fig_pairv}. 
This is demonstrated by Fig.~\ref{fig_pairv_nsr} where $h$ determined
directly is compared with values obtained from the non-linear scaling
relation. 

The value of $h$ fluctuates a lot at late epochs, we mentioned in the previous
section that the slope of the correlation function gives a more stable
estimate of $h$.  
Due to this, some of the late epochs used in Fig.~\ref{fig_corr}
have not been plotted in Fig.~\ref{fig_pairv}. 
The key feature in this regime is that $h < 1$ and it does not show any sign
of heading towards the stable clustering limit of $h=1$. 
This is consistent with the results from the slope of the correlation
function.  

It is important to understand what $h < 1$ means in terms of dynamics.  
One possibility is that the haloes are evaporating.  
This can happen if two-body relaxation is important, but we have used
$\epsilon \geq 0.2$ grid lengths in all our simulations and number density of
particles in most haloes is very high.  
Hence, two-body scattering should not be important.  
A poor integrator for the equation of motion can also lead to the evaporation
of haloes.  
We have tested the integrator in three-body problems and with
highly eccentric binary orbits to ensure that the problem does not lie with
incorrect evolution of trajectories.

In order to probe the dynamics further, we also study non-radial motions. 
Analogous to $h$, the radial component of the pair velocity scaled by Hubble
velocity, we define the transverse pair velocity function as:
\begin{equation}
g(r,a) = \left| \frac{\langle{{v_p}_x y_p - {v_p}_y x_p}\rangle}{H r^2} \right|
\label{geqn}
\end{equation}
Here the angle brackets denote averaging over all pairs with separation $r$,
${\mathbf v}_{p}$ is the pair velocity (peculiar velocity) and $ {\mathbf
  r}_{p}$ is the separation of the pair of particles.
$H$ is the Hubble parameter.
The equivalent object in three dimensions is:
\begin{equation}
g(r,a) =\left| {\mathbf g} \right| = \frac{\left| {\mathbf v} ~ \times ~
  {\mathbf r}   \right|}{H ~ r^2}
\end{equation}
In two dimensions the ``cross'' product of two vectors is just a number.
Fig.~\ref{fig_pairvt} shows $|g|$ as a function of $\bar\xi$.
We have plotted the magnitude of $g(r,a)$ as its value oscillates
about zero at large-scales. 
This curve is plotted for only one epoch but the relative value of $|g|$
compared to $h$ shows that this is a significant component. 
Note that we require coherent transverse motions in order to detect anything
here as random transverse motions will cancel out in the sum.  
We do not find any systematic excess in pairwise transverse velocity
dispersion as compared to the pairwise collinear velocity dispersion.
In dynamical equilibrium these should have the same magnitude in two
dimensions and at $\bar\xi \geq 10$ we find this to be true. 


\begin{figure}
\includegraphics[width=3.0truein]{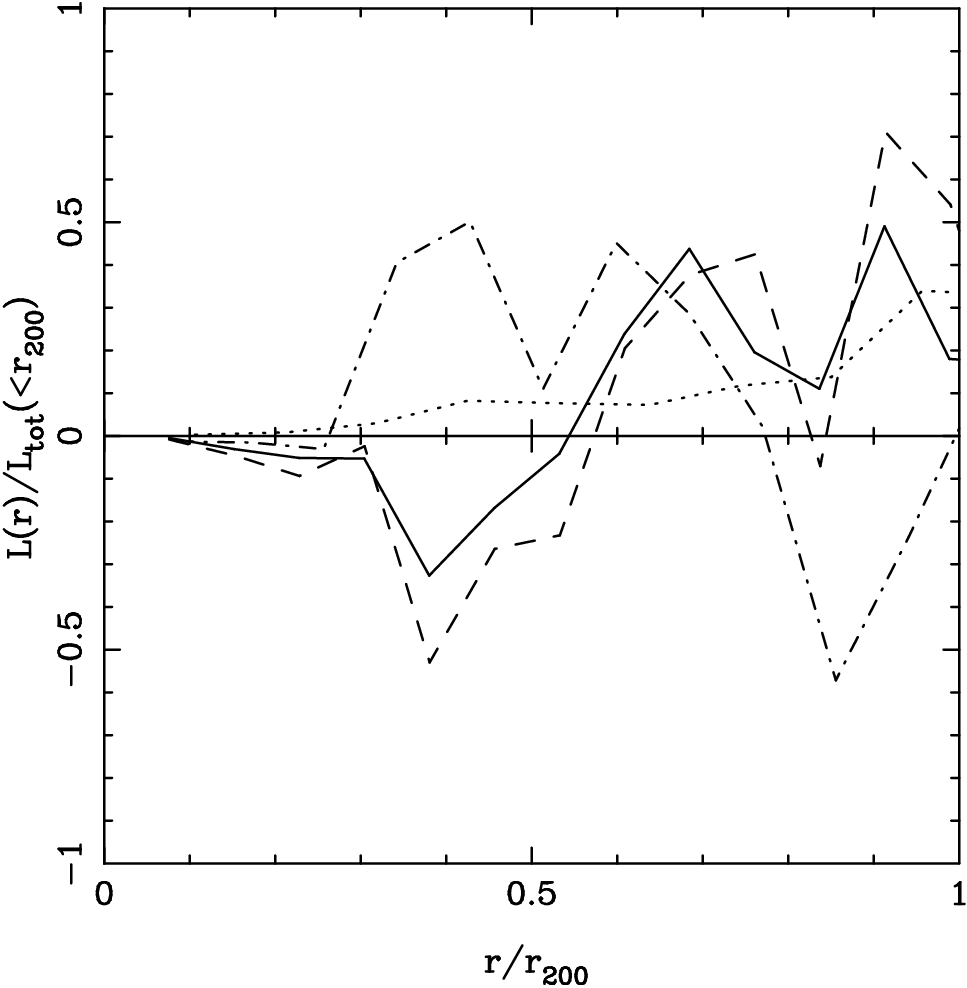} 
\includegraphics[width=3.0truein]{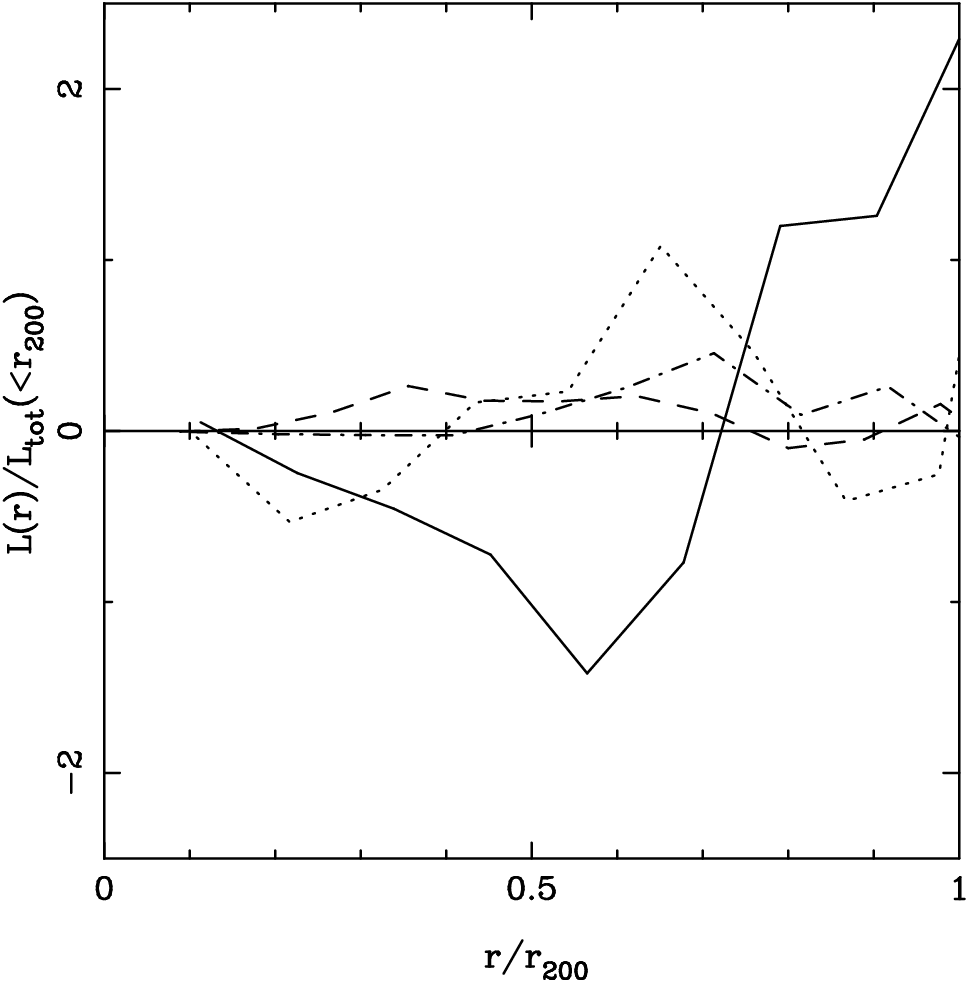}
\caption{In this figure, we plot circularly averaged angular momentum
  of particles in a halo at a distance $r$ from the centre of mass
  of the halo for some haloes.  Distance $r$ is scaled by $r_{200}$ and the
  contribution of particles at a distance $r$ to the angular momentum is
  scaled by the total angular momentum contained within $r_{200}$.
  The upper panel is for index $n=-0.4$
  and the lower for $n=1.0$.  Different line styles here correspond
  to different haloes.} 
\label{fig_angmom}
\end{figure}


To investigate this further, we have studied motions in individual haloes by
plotting the angular momentum profile as a function of distance from the
centre of the halo. 
There are no systematic features of shape worth commenting except that
for most haloes there are annuli with coherent rotation.  
Note that this does not imply a coherent rotation for the halo as the
direction of rotation reverses a few times.
Fig.~\ref{fig_angmom} shows circularly averaged angular momentum of
particles in a halo within a distance $r$ from the centre of mass of
the halo for some (randomly selected) haloes.  
The upper panel is for index $n=-0.4$ and the lower for $n=1.0$.  
Clearly, all the haloes in both panels have annuli with coherent
rotation.  
In some cases, angular momentum contributed by adjacent annuli is in opposite
direction. 
The net angular momentum in these haloes contributes a small fraction of the
kinetic energy of particles in these haloes. 
A discussion of the selection criteria used to identify haloes is given in the
next sub-section.  

It is important to understand the origin of this apparent coherent rotation in
massive haloes.  
Comparison of $g-\bar\xi$ plot (Fig.~\ref{fig_pairvt}) for $n=-0.4$ and
$n=1$ offers us a clue, transverse motions are much stronger for $n=1$ as
compared to the other model. 
Collisions between sub-structure falling into haloes in the early
phase can generate such transverse motions
\cite{2004astro.ph..8429B}, and there is certainly more substructure
in the $n=1$ model. 
We put forward the hypothesis that collisions between substructure are
responsible for generation of coherent transverse motions. 
We are testing this hypothesis in a series of numerical simulations.  
Results of these studies will be reported elsewhere.


\begin{figure}
\includegraphics[width=3.2truein]{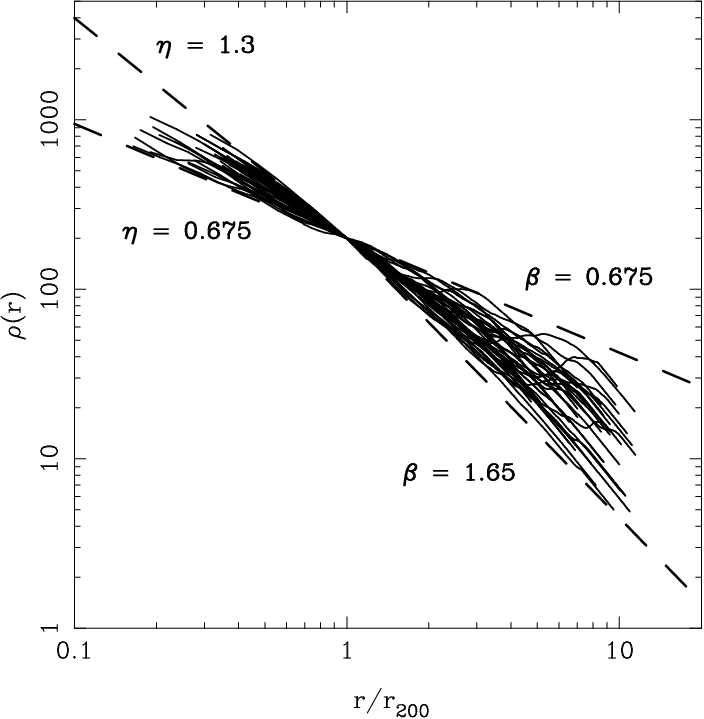} 
\includegraphics[width=3.2truein]{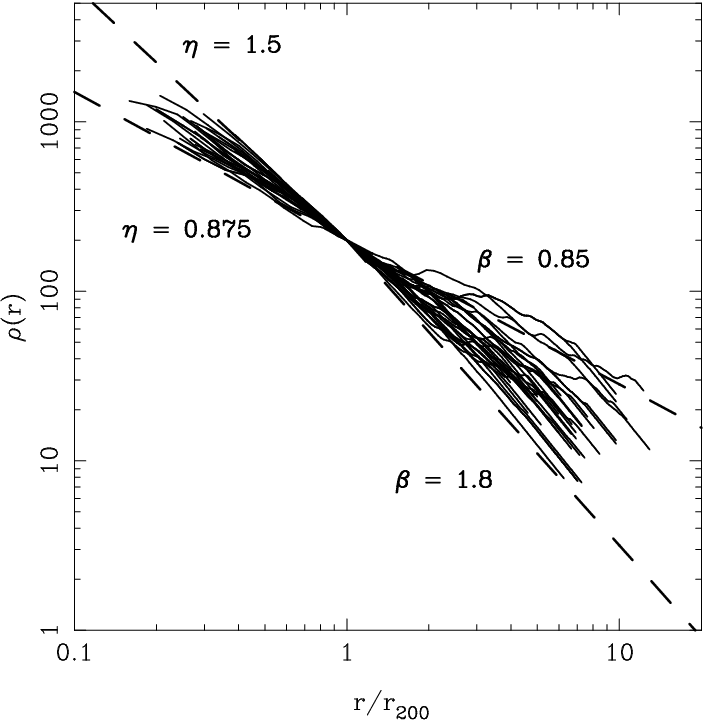}
\caption{Spherically averaged density profiles of a
  number of spherically symmetric virialised haloes from simulations
  with indices $n=-0.4$ (upper panel) and $n=1.0$ (lower
  panel). Density is plotted as a function of $r/r_{200}$. The four dashed
  lines in each panel mark 
  the approximate extremes of inner ($\eta$) and outer slopes
  ($\beta$) for the density profiles. 
 The smallest haloes have more than $10^3$ particles inside the virial radius,
  and the largest haloes have more than $10^4$ particles. 
All the haloes in the figure
  were taken from TreePM simulations with $2048^2$ particles.} 
\label{fig_denprofiles} 
\end{figure}


\subsection{Density Profiles of Massive Haloes}

Density profiles of massive haloes and non-linear scaling relations
are closely related.  
Density profiles of very massive haloes are easier to study as these
extend over a large region, large compared to the softening length
which is the smallest scale we can resolve in our simulations. 
We identified haloes in our simulations using the method of density
peaks \cite{1997MNRAS.286..865T}.
In this method the density field is smoothed at a large length scale
$R=5$ grid lengths and peaks in the smoothed density field are identified.   
Particles within a distance $R$ of a peak are selected and we find
the centre of mass for these particles.
We now select a smaller set of particles out of these, those that are
within $R-\Delta R$ from the centre of mass and compute the centre of
mass. 
These iterations continue till we are left with only a small number of
particles, say about $100$.  
The final centre of mass is taken to be the centre of the halo and the density
profiles are plotted as a function of distance from this centre. 
We apply two further criteria:
\begin{itemize}
\item
Central density contrast of the halo must be very large, $\delta \geq
500$. 
\item
Haloes and their neighbourhood should be smooth and there should be no
major merger going on.
To implement this in an objective manner we compute the ratio $\chi =
\langle \cos^2 \phi \rangle / \langle \sin^2 \phi \rangle $ where
average is over all particles within the central region of the halo
and $\phi$ is the position angle of a particle from the centre of
mass. 
We require $0.9 \geq \chi \geq 1.1$ for a halo to be used for density
profiles. 
\end{itemize}
These two criteria reject about $75\%$ of all the haloes with rejection being
more frequent for models with a larger index $n$.

Fig.~\ref{fig_denprofiles} shows the density profiles of haloes that
satisfy these criteria in our simulations.  
Density profiles of a large number of haloes are plotted, for each
halo we have drawn density $\rho$ as a function of $r/r_{200}$, where
$r_{200}$ is the scale at which the density is $200$ times the
average. 
Average density is unity in these simulations.
It is clear that there is a large scatter in density profiles.  
The inner slope varies from $0.4$ to $1.4$ for $n=-0.4$ whereas it is in
the range $0.6$ to $1.6$ for $n=1$. 
The inner slope $\eta$ is obtained by fitting a power law that
passes through $\rho=200$ and $r/r_{200} = 1.0$ and only points inside
of this radius are used. 
The distribution of the inner slope of density profiles is very
broad, as is obvious from the range.  
The median $\eta$ is approximately $0.9$ for $n=-0.4$ and $1.1$
for $n=1.0$. 
Self-similar, spherical collapse predicts an inner slope of $\eta=0.89$ and
$\eta=1.2$ for $n=-0.4$ and $n=1$, respectively (eqn.(2)).  
These values are well within the scatter of inner slopes. 
Clearly, the answer lies in between the prediction for the self-similar
collapse and a universal slope.

It is difficult to explain the large scatter seen in density profiles.  
The origin of scatter can lie in environmental dependence as well as in
different formation histories for haloes.  
We can, however, address the issue of the small systematic shift in median
slope as compared to that expected from stable clustering. 

Transverse motions can affect the density profile of massive haloes in
a significant manner \cite{2000ApJ...538..517S}. 
Such effects will be stronger if the transverse motions are stronger,
and as is obvious from the Fig.~\ref{fig_pairvt} transverse motions
are stronger for larger $n$. 
Rotation is playing a more important role in models with more
substructure and is leading to flatter density profiles.  
This reduces the difference between density profiles of massive haloes
for these models.
This, we believe, is a common explanation for the non-linear scaling
relations and density profiles.


\begin{figure}
\includegraphics[width=3.2truein]{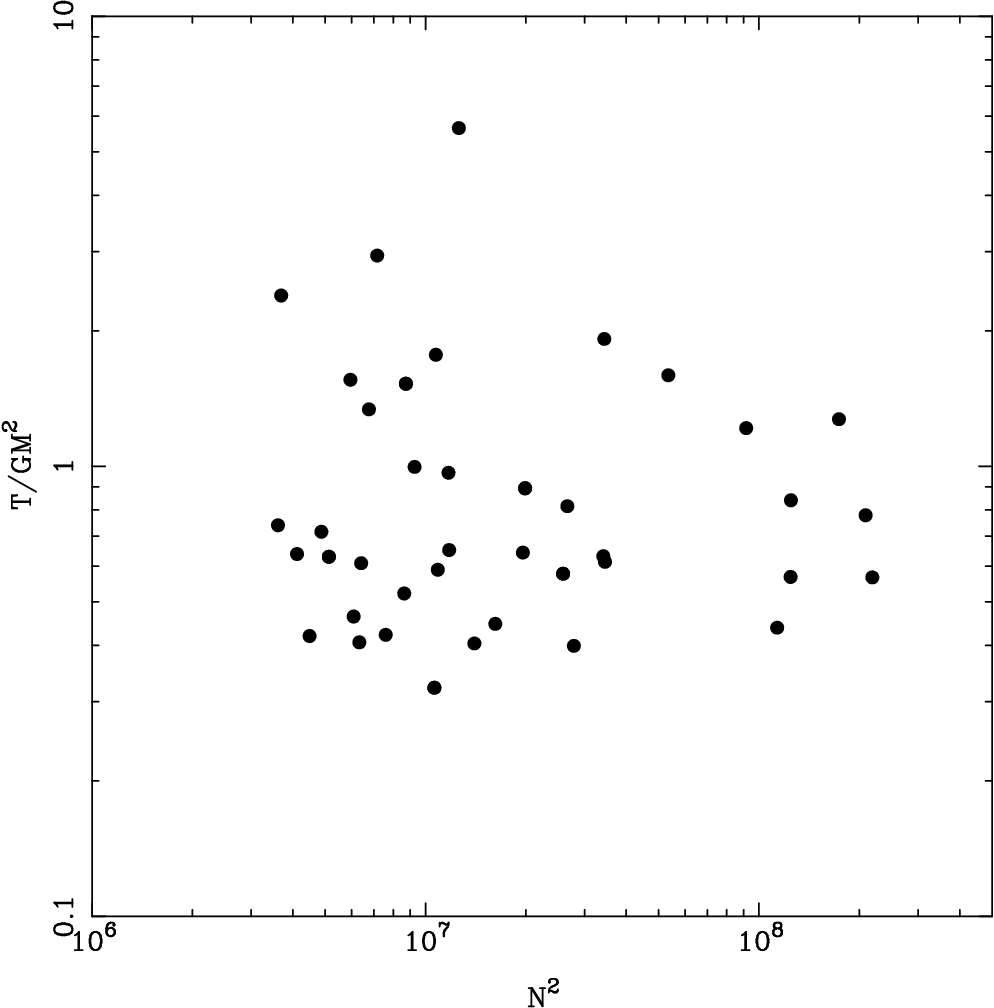} 
\caption{The average kinetic energy $T$ of massive haloes is scaled by $G~M^2$
  and plotted against $M^2$ on a log-log scale.  Each point on this graph
  represents one halo, we  have used the set of haloes used for studying
  density profiles.  This plot shows that there is large scatter about the
  average relation $T \propto M^2$.}
\label{fig_vir}
\end{figure}


\subsection{Virial Equilibrium}

The stable clustering limit requires formation of objects that are in
dynamical equilibrium.  
Following Peebles (1980), we compute the second time derivative of the moment
of inertia and set it to zero for a cluster of particles in dynamical
equilibrium.  
This gives us the condition to be satisfied for dynamical equilibrium.  
We get:
\begin{equation}
\sum\limits_i \frac{1}{2} m_i {\dot{\mathbf r}}_i^2 = \frac{1}{2} \sum\limits_i
\sum\limits_{j \neq i} G m_i m_j 
\end{equation}
where the sums are over all particles and dots denote time derivatives.  
Particles are located at ${\mathbf r}_i$ and have velocities ${\dot{\mathbf
    r}}_i$ and $m_i$ is the mass of the $i$th particle.  
Thus the kinetic energy depends {\it only on the total mass of the halo.}
If the massive haloes that we are studying here are in virial equilibrium then
the kinetic energy $T$ for these haloes should be proportional to $M^2$. 
For each halo used for density profiles, all particles within the radius
$r_{200}$ were used to plot Fig.~\ref{fig_vir}.
It is clear that these quantities are correlated even though there is a large 
scatter about the average relation.  
Thus we may conclude that the massive haloes are close to dynamical
equilibrium.


\section{Conclusions}

The basic motivation for this work was to improve our understanding of
non-linear scaling relations for two dimensional gravitational
clustering.  
Previous studies \cite{1998ApJ...495...25B,1998MNRAS.293L..68M}
established the behaviour for the quasi-linear regime, but could not probe the 
nature of the asymptotic regime.   
We used the 2D TreePM code \cite{2004astro.ph..6009R} which allows us to study
the non-linear regime in far greater detail, as compared to the PM code used
in earlier studies. 
The results of our study can be summarised as follows:
\begin{itemize}
\item
The evolution of the correlation function for power-law models is scale
invariant up to the highest non-linearities reached in our
simulations.   
\item
We reproduce the results of earlier studies in the quasi-linear
regime. 
We confirm that the slope of the non-linear scaling relation in this
regime is close to $2$, as predicted on the basis of the infall dominated
growth model \cite{1996MNRAS.278L..29P}.
\item
We do not find any difference in the non-linear scaling relation for
different power law models.  
Thus the information about initial conditions is retained even in the
extremely non-linear regime.
\item
The stable clustering limit ($h \rightarrow 1$ as $\bar\xi \rightarrow
\infty$) is not reached.  
(Though it is generally claimed that stable clustering with $h\to 1$ is a
{\sl natural} asymptotic state, it has been shown that the Davis-Peebles
scale-invariant solution \cite{1977ApJS...34..425D} and the hierarchical model
for the three-point function are inconsistent with the standard
stable-clustering picture \cite{2000ApJ...531...17K}. 
This, of course, is for gravitational collapse in three dimensions.)
This can be concluded from the slope of the correlation function, pair
velocity, as well as density profiles of massive haloes.
\item
Pair velocities in the asymptotic regime are smaller than expected
from the stable clustering model. 
This does not imply that massive haloes are evaporating. 
\item
The transverse component of pair velocity is more significant for
models with a larger spectral index, i.e., for models with more
substructure. 
This suggests that gravitational collisions between substructure orbiting
within haloes are responsible for generating coherent transverse motions.
\item
We find that there is no universal density profile for massive haloes
in two dimensional gravitational clustering. 
There is a large scatter in inner as well as outer slopes of density
profiles. 
Median value of the inner slope is different for different models,
though the difference is much smaller than the scatter in values. 
Tests show that these clusters are close to dynamical equilibrium. 
\item
The difference in the median inner slope for different models is smaller than
expected from self-similar spherical collapse.
We argue that the difference in significance of transverse motions
for different models is responsible for this change.
\end{itemize}

Many of the conclusions listed here have been checked independently in large
PM simulations ($4096^2$ particles) up to $\bar\xi \leq 100$. 

Further work is  required to test our hypothesis that
gravitational collisions between substructure are responsible for
generating coherent transverse motions. 
We are carrying out a detailed study using numerical simulations where
we tune the amount of substructure to test this proposal. 

Considerable work has been done to study the non-linear scaling
relations in three dimensions
\cite{1991ApJ...374L...1H,1994MNRAS.271..976N,1994MNRAS.267.1020P,1995MNRAS.276L..25J,1996MNRAS.280L..19P,1996MNRAS.278L..29P,1996ApJ...466..604P,1998ApJ...495...25B,2003MNRAS.341.1311S}.  
We intend to study these again in light of insight developed using
tests in clustering in two dimensions. 
It has been shown that the scaling relations for gravitational clustering in
three dimensions depend on the initial spectrum of fluctuations
\cite{2003MNRAS.341.1311S}. 
It is important to understand the origin of this index dependence and test
whether substructure plays a role in this. 
The transverse pair velocity function has been introduced in eqn.(\ref{geqn}),
we are studying the three dimensional equivalent of this quantity in order to
probe the role of transverse motions in gravitational clustering in the full
three dimensional problem. 
These studies should help us understand the non-linear scaling relations
better.


\section*{Acknowledgements}

The numerical work reported here was done using the Beowulf at the
Harish-Chandra Research Institute (http://cluster.mri.ernet.in).  
This research has made use of NASA's Astrophysics Data System. 



\label{lastpage}

\end{document}